\documentclass[letterpaper,12pt]{article}
\usepackage{tabularx} 
\usepackage{amsmath}  
\usepackage{graphicx} 
\graphicspath{ {./Figures/} }
\usepackage{float} 
\usepackage[margin=1in,letterpaper]{geometry} 
\usepackage{epstopdf}
\usepackage{cite} 
\usepackage[final]{hyperref} 
\hypersetup{
	colorlinks=true,       
	linkcolor=blue,        
	citecolor=blue,        
	filecolor=magenta,     
	urlcolor=blue         
}

\begin{document}

\title{\textbf{Taking the Road Less Traveled: Solving the One-Dimensional Quantum Oscillator using the Parabolic-Cylinder Equation}
\author{M\'{a}t\'{e} Garai$^{1}$, Douglas A. Barlow $^1$\\
\small $^{1}$ Department of Physics, The University of the South, Sewanee, TN 37383}}

\date{}
\maketitle

\begin{abstract}
The single well 1D harmonic oscillator is one of the most fundamental and commonly solved problems in quantum mechanics. Traditionally, in most introductory quantum mechanics textbooks, it is solved using either a power series method, which ultimately leads to the Hermite polynomials, or by ladder operators methods. We show here that, by employing one straightforward variable transformation, this problem can be solved, and the resulting state functions can be given in terms of parabolic cylinder functions. Additionally, the same approach can be used to solve the Schr\"{o}dinger equation for the 1D harmonic oscillator in a uniform electric field. In this case, the process yields two possible solutions. One is the well-known result where the 1D oscillator eigenvalues are reduced by a frequency-dependent term, which can have any positive value. The other is where the field term is restricted to be an integer and the eigenvalues are in the same form as for the field-free case. Additionally, we show how the results can be used to create a harmonic approximation for the bound states of a Lennard-Jones potential.
\end{abstract}


\clearpage

\section{Introduction}
The single well 1D harmonic oscillator is one of the most fundamental and commonly solved problems in undergraduate quantum mechanics courses. Traditionally, in most introductory quantum mechanics textbooks, the governing differential equation is solved analytically using a power series method, which ultimately leads to the Hermite polynomials, or algebraically using ladder operators. In fact, one or both of these methods were the exclusive approach taken in the more than thirty texts we reviewed for this study where the solution for the 1D quantum oscillator was given and discussed \cite{Abers, Auletta, Band, Beard, Bohm, Blumel, Cohen, Dirac, Eisberg, Fayer, Fong, French, Gas, Griffiths, Gott, Houston, Johnson, Kon, Lawden, Liboff, Mandel, Merz,  Morrison2, Morrison, Park, Pauling, Peebles, Phillips, Powell, Rae, Rob, Roj, Sak, Saxon, Scher, Shan, Strauss, Townsend, Zet}. However, we demonstrate here a less laborious route to the complete solution for this problem. By employing one straightforward variable transformation, the governing equation can be written in the form of a parabolic cylinder equation, and the resulting state functions are thus given in terms of the parabolic cylinder functions. In the list of texts mentioned above, only Merzbacher \cite{Merz} utilized a parabolic cylinder equation solution for any oscillator problem, and this is the double harmonic oscillator, not the single well 1D oscillator that we consider here. 

Additionally, a similar approach can be used to solve the Schr\"{o}dinger equation for the 1D harmonic oscillator in a uniform electric field. In this case, the resulting parabolic cylinder equation reveals the possibility of two solutions. One is the well-known result where the 1D oscillator eigenvalues are reduced by a frequency-dependent term that can have any positive value \cite{Johnson}. The other is where the field term is restricted to be an integer, and the eigenvalues are more like those for the field-free case.

In the next section, steps are given for arriving at the solution for the 1D harmonic oscillator in terms of parabolic cylinder functions. In the subsequent section, the case where the oscillator is within a uniform electric field is considered. Again, steps leading to the solution in terms of parabolic cylinder functions are given. Finally, we demonstrate how the results for the oscillator in a uniform electric field can be used to approximate the bound states of a Lennard-Jones potential.

 \section{Solutions to the Quantum Harmonic Oscillator}
We consider the classical potential energy, $V$, for the one dimensional (1D) harmonic oscillator as, $V=\frac{1}{2}\mu\omega^2x^2$, where $\mu$ is reduced mass, $k$ is the force constant and $x$ is position. It is useful to write the force constant in terms of the oscillator's natural frequency $\omega$ as $\omega=\sqrt{k/\mu}$. With this substitution, the full Hamiltonian is written as
\setlength{\jot}{10pt}
\begin{equation}
\hat{H} = -\frac{\hbar^2}{2\mu}\frac{d^2}{dx^2}+\frac{1}{2}m\omega^2x^2~.
\label{eq:1}
\end{equation}
Therefore, the time-independent Schr\"{o}dinger equation for the 1D harmonic oscillator is:
\begin{equation}
-\frac{\hbar^2}{2\mu}\frac{d^2}{dx^2}\psi+\frac{1}{2}\mu\omega^2x^2\psi=E\psi~.
\label{eq:2}
\end{equation}
After a bit of manipulation this becomes
\begin{equation}
\frac{\hbar}{2\mu\omega}\frac{d^2}{dx^2}\psi+\left(\frac{E}{\hbar\omega}-\frac{\mu\omega}{2\hbar}x^2\right)\psi=0~.
\label{eq:3}
\end{equation}
The parabolic cylinder differential equation is of the form \cite{Bateman}:
\begin{equation}
\frac{d^2y}{dz^2}+\left(n+\frac{1}{2}-\frac{1}{4}z^2\right)y=0~,
\label{eq:4}
\end{equation}
where $n$ is an integer. Eq. (\ref{eq:3}) can be transformed into the form of Eq. (\ref{eq:4}) with the aid of the following substitution:
\begin{equation}
z=\sqrt{\frac{2\mu\omega}{\hbar}}x~.
\label{eq:5}
\end{equation}
From Eq. (\ref{eq:5}) one finds that
\begin{equation}
\frac{d^2 \psi}{dx^2} = \frac{2 \omega \mu}{\hbar} \frac{d^2 \psi}{dz^2}~.
\label{eq:6}
\end{equation}
Using Eqs (\ref{eq:6}) and (\ref{eq:5}) in Eq. (\ref{eq:3}) leads to
\begin{equation}
\frac{d^2\psi}{dz^2}+\left(\frac{E}{\hbar\omega}-\frac{z^2}{4}\right)\psi=0~.
\label{eq:7}
\end{equation}
As we are expecting quantized energy levels we set $E=E_n$ and let
\begin{equation}
E_n = \hbar \omega (n +1/2)~.
\label{eq:8}
\end{equation}
With this substitution, Eq. (\ref{eq:7}) is in the form of the parabolic cylinder equation so that the state functions $\psi_n$ are given by the complete orthogonal set of solutions for Eq. (\ref{eq:7}), the \textit{parabolic cylinder functions}, $\{D_n (z)\}$ for $n = 0, 1, 2, \dots$. The state functions for the 1D oscillator are then: 
\begin{equation}
\psi_n = N_n\,D_{n}\left(\sqrt{\frac{2\mu\omega}{\hbar}}x\right)~,
\label{eq:9}
\end{equation}
where $N_n$ is a normalization constant. 

Not surprisingly, the functions, $D_n$ can be written in terms of the Hermite polynomials, $H_n$ as \cite{Bateman}
\begin{equation}
D_{n}(z)= \frac{1}{(\sqrt{2})^n} e^{-z^2/4}H_{n}\left(\frac{z}{\sqrt{2}}\right)~,
\label{eq:10}
\end{equation}
which, due to the nature of the Hermite polynomials, restricts $n$ to non-negative integers. Using the Rodrigues formula for the Hermite polynomials, \cite{Bell}, we can write an \textit{effective Rodrigues formula} for the parabolic cylinder functions as
\begin{equation}
D_n (z) = (-1)^n e^{-\frac{z^2}{4}} \frac{d^n}{d z^n}e^{-\frac{z^2}{2}}~.
\label{eq:11}
\end{equation}
The first six parabolic cylinder functions are listed in Table 1. A few of these functions are plotted in Figure 1.
\begin{table}[h]
\begin{center}
\caption{\small{The first six parabolic cylinder functions $D_{n}(z)$ }}
\begin{tabular}{ll}
\hline \hline\vspace{2mm}
{n} & ~~~~~~~~~~~{$D_{n}(z)$ }\\
0& ~~~~~~~~~~~~$e^{-z^2/4}$\\
1& ~~~~~~~~~~~~$e^{-z^2/4}z$\\
2& ~~~~~~~~~~~~$-e^{-z^2/4} (1 - z^2)$\\
3& ~~~~~~~~~~~~$e^{-z^2/4} z (z^2 -3)$\\
4& ~~~~~~~~~~~~$-e^{-z^2/4} (6 z^2 -3 - z^4)$\\
5& ~~~~~~~~~~~~$e^{-z^2/4}z (15 - 10 z^2 + z^4)$\\
\hline
\end{tabular}
\end{center}
\end{table}

\begin{figure}[h]\label{fig:parabolicplot}
  \centering
  \includegraphics[width=0.75\linewidth]{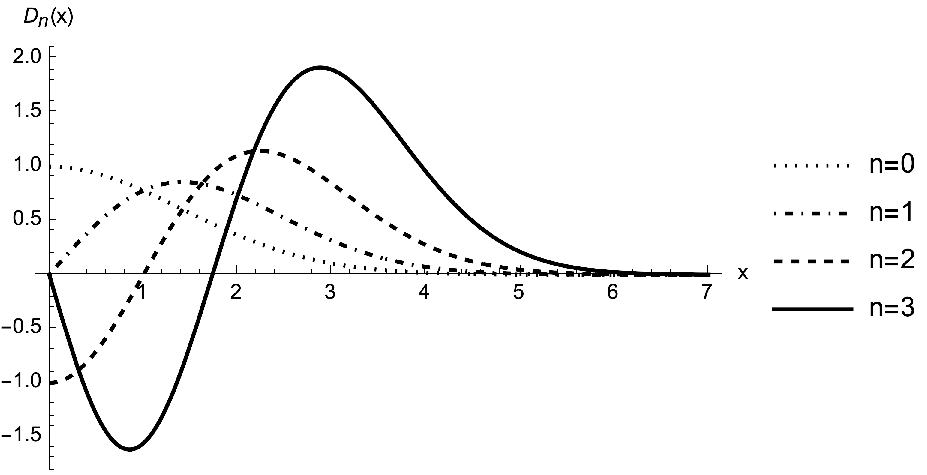}
\caption{The first four parabolic cylinder functions.} 
\end{figure}

The normalization constant can now be found using the following identity from Reference \cite{Lebedev}:
\begin{equation}
\int_{-\infty}^{\infty}e^{-x^2} {H_n}^2 (x) dx = 2^n n! \sqrt{\pi}~.
\label{eq:12}
\end{equation}
Using Eqs. (\ref{eq:9}) and (\ref{eq:10}) we can write that
\begin{equation}
\langle \Psi_n | \Psi_n \rangle = \frac{{N_n}^2}{2^n} \int_{-\infty}^{\infty} e^{-z^2 /2} {H_n}^2 \left(\frac{z}{\sqrt{2}}\right) dz = 1~.
\label{eq:13}
\end{equation}
Letting $z$ be given by Eq. (\ref{eq:5}) in Eq. (\ref{eq:13}), and using Eq. (\ref{eq:12}) to evaluate the integral, leads to
\begin{equation}
N_n =\frac{1}{\sqrt{n!}}\left(\frac{\mu\omega}{\hbar \pi}\right)^{1/4}~.
\label{eq:14}
\end{equation}
Therefore, the complete set of state functions for the quantum harmonic oscillator is expressed in terms of parabolic cylinder functions as:
\begin{equation}
\psi_n(x)=\frac{1}{\sqrt{n!}}\left(\frac{\mu\omega}{\hbar \pi}\right)^{1/4} D_{n}\left(\sqrt{\frac{2\mu\omega}{\hbar}}x\right)~.
\label{eq:15}
\end{equation}
Using these state functions, the well-known energy eigenvalues and position expectation value for this problem can be computed. That is,
\begin{equation}
\hat{H} | \Psi_n \rangle = \hbar \omega (n +1/2)\Psi_n~,
\label{eq:16}
\end{equation}
and
\begin{equation}
\langle \Psi_n |x| \Psi_n \rangle = 0~.
\label{eq:17}
\end{equation}

 \section{The Harmonic Oscillator in a Uniform Electric Field:}
The method of solution using parabolic cylinder functions is also useful when the oscillator Hamiltonian involves an additional first-order term. The physical interpretation of this case might involve an oscillator placed in a uniform electric field of magnitude $\mathcal{E}$. In this case, the Hamiltonian is of the form:
\begin{equation}
\hat{H}=-\frac{\hbar^2}{2\mu}\frac{d^2}{dx^2}+\left(\frac{1}{2}\mu\omega^2x^2+q\mathcal{E}x\right)~.
\label{eq:18}
\end{equation}
The 1D time-independent Schr\"{o}dinger equation is then written as:
\begin{equation}
-\frac{\hbar^2}{2\mu}\frac{d^2\psi}{dx^2}+\left(\frac{1}{2}\mu\omega^2x^2+q\mathcal{E}x-E\right)\psi=0~.
\label{eq:19}
\end{equation}
Using the substitution from Eq. (\ref{eq:5}), which we now label as $u = \sqrt{(2\mu\omega/\hbar})x$, in Eq. (\ref{eq:19}) yields
\begin{equation}
-\hbar\omega\frac{d^2\psi}{du^2}+\left(\frac{\hbar\omega}{4}u^2+q\mathcal{E}\sqrt{\frac{\hbar}{2\omega \mu}}u-E\right)=0~.
\label{eq:20}
\end{equation}
Dividing this by $-\hbar\omega$ we arrive at the following:
\begin{equation}
\frac{d^2\psi}{du^2}+\left(\frac{E}{\hbar\omega}-\frac{u^2}{4}-\frac{q \mathcal{E}}{\sqrt{2\mu\hbar \omega^3}}u\right)=0~.
\label{eq:21}
\end{equation}
It is useful now to define the following constant:
\begin{equation}
\gamma = \frac{q \mathcal{E}}{\sqrt{2\mu\hbar \omega^3}}~.
\label{eq:22}
\end{equation}
Using this in Eq. (\ref{eq:21}) gives
\begin{equation}
\frac{d^2\psi}{du^2}+\left(\frac{E}{\hbar\omega}-\frac{u^2}{4}-\gamma u\right)=0~.
\label{eq:23}
\end{equation}
Now, we apply the additional variable transformation, $z=u + 2\gamma$, so that Eq. (\ref{eq:23}) is converted into the form of the parabolic cylinder differential equation
\begin{equation}
\frac{d^2\psi}{dz^2}+\left(\frac{E}{\hbar\omega}-\frac{z^2}{4}+\gamma^2\right)\psi=0~.
\label{eq:24}
\end{equation}
We are presented with two choices on writing the solution for Eq. (\ref{eq:24}). First, consider the solution where
\begin{equation}
E_n=\hbar\omega(n+1/2-\gamma^2)~.
\label{eq:25}
\end{equation}
Here we see that the energy levels of the harmonic oscillator will be shifted by the amount $\gamma^2$ due to the influence of the electric field. This is, of course, the classic result that can also be obtained via perturbation methods \cite{Johnson}. The normalized state functions in the case, which we label as $\Psi_{n, \gamma^2}$, are
\begin{equation}
\Psi_{n, \gamma^2}(x) = N_{n} D_{n}\left(\sqrt{\frac{2\mu\omega}{\hbar}} x + 2\gamma\right)~.
\label{eq:26}
\end{equation}
Following the steps in the previous section, one can confirm that the normalization constant in this case is again given by Eq. (\ref{eq:14}). One can show that eq. (\ref{eq:26}) does indeed give the eigenfunctions for the eigenvalues given by Eq. (\ref{eq:25}).

However, it is interesting to consider a second solution for this problem afforded by Eq. (\ref{eq:24}). Here, we let 
\begin{equation}
E_m = \hbar\omega(m +1/2)~,
\label{eq:27}
\end{equation}
where $m$ is an integer. Inserting this into Eq. (\ref{eq:24}) gives
\begin{equation}
\frac{d^2\psi}{dz^2}+\left(m + \frac{1}{2} -\frac{z^2}{4}+\gamma^2\right)\psi=0~.
\label{eq:28}
\end{equation}
So, it must be that $m + \gamma^2 \ge 0$. We define the integer $k$ where $k = m +\gamma^2$ and $k = 1, 2, 3, \dots$ and thus $\gamma^2 = 1, 2, 3, \dots$ and then $m = -\gamma^2,~~ \gamma^2 - 1, ~~\gamma^2 - 2 \dots$ . With this substitution, Eq. (\ref{eq:28}) is in the form of Eq. (\ref{eq:4}) and thus has as solutions the following state functions, which we label as $\Psi_{m +\gamma^2}$,
\begin{equation}
\Psi_{m +\gamma^2} = N_n D_{m +\gamma^2}\left(\sqrt{\frac{2\mu\omega}{\hbar}} x + 2\gamma\right)~,
\label{eq:29}
\end{equation}
where $N_n$ is again given by Eq. (\ref{eq:14}). Thus for any $m$ in the range of $-\gamma^2$ to any greater integer, Eq. (\ref{eq:29}) gives the eigenfunctions for the eigenvalues of Eq. (\ref{eq:27}) when $\gamma^2$ is a positive integer. That is,
\begin{equation}
\hat{H}|\Psi_{m +\gamma^2} \rangle = \hbar\omega(m+1/2)|\Psi_{m +\gamma^2} \rangle~,
\label{eq:30}
\end{equation}
for any $\gamma^2 = 1, 2, 3, \dots$. So, it can be seen that the choice for the energy given by Eq. (\ref{eq:27}) filters out the values for $E_n$ in in Eq. (\ref{eq:25}) for cases where $\gamma^2$ is a positive integer.

\begin{figure}[h]\label{fig:parabolicplot}
  \centering
  \includegraphics[width=0.75\linewidth]{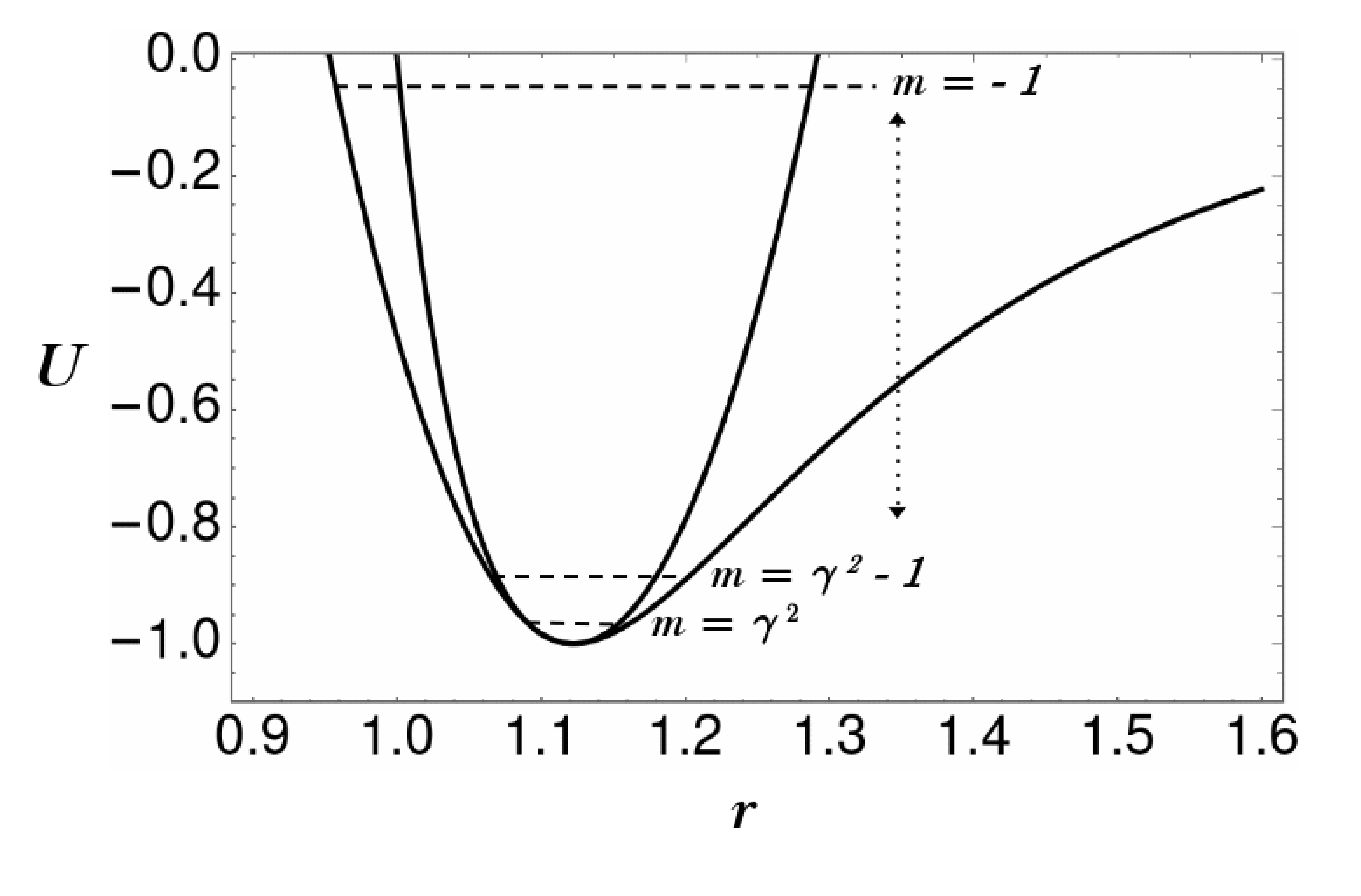}
\caption{Curves for a L-J potential well and its harmonic oscillator approximation with some of the energy levels denoted. $\epsilon$ and $\sigma$ are set to 1 and $k = 70$ in arbitrary units. } 
\end{figure}

\section{Discussion and Application}

The solution for the harmonic oscillator described in the previous section, where the eigenvalues are given by Eq. (\ref{eq:27}), gives nothing new as these solutions are a subset of those for the more general situation where $\gamma$ can have any positive value. However, this solution does awaken us to the fact that this model can be used when the values of $\gamma^2$ are restricted to the positive integers.

It is important to note that the effect of the electric field on the oscillator is to introduce negative energies to the eigenvalue spectrum. With this in mind, we can use the solution for the case where $\gamma^2$ is a positive integer to create an oscillator model for some other type of bound state potential for which the Schr\"{o}dinger equation might be difficult to solve. As an example, we consider the classic Lennard-Jones (L-J) potential, $U(r)$, which is often used to describe bonding between two noble element atoms separated by a distance $r$:
\begin{equation}
U(r) = 4\epsilon\left[(\sigma/r)^{12} - (\sigma/r)^6 \right]~.
\label{eq:31}
\end{equation}
Here $\epsilon$ and $\sigma$ are the L-J parameters. We demonstrate here, how to use the results of the previous section to approximate the bound states of this potential as a harmonic oscillator. This process is facilitated by the fact that $\gamma^2$ can be restricted to integer values. The curve for a L-J potential, and an approximating harmonic oscillator curve are depicted in Figure 2.

The well known energy minimum for the L-J potential is simply $-\epsilon$. To relate this to the harmoinc oscillator in the electric field we find the minimum energy for the potential in the Hamiltonian of Eq. (\ref{eq:18}). One finds by differentiation, and setting the result to zero, that this occurs when $x = -q\mathcal{E}/(\mu\omega^2)$. Inserting the value for $x$ back into the oscillator potential yields the energy minimum $E_{min}$:
\begin{equation}
E_{min} = -\frac{(q\mathcal{E})^2}{\mu \omega^2}~.
\label{eq:32}
\end{equation} 
Setting this equal to $\epsilon$ and using Eq. (\ref{eq:22}) we find that
\begin{equation}
\hbar \omega = \frac{\epsilon}{\gamma^2}~.
\label{eq:33}
\end{equation}
Now, let $\gamma^2$ be the total number of bound states and $m = -\gamma^2, -\gamma^2 + 1, -\gamma^2 + 2, \dots -1$ so that the energy levels of our approximating oscillator become
\begin{equation}
E_m = \frac{\epsilon}{\gamma^2} \left(m + 1/2 \right)~.
\label{eq:34}
\end{equation}
The minimum of the L-J potential can be shown to occur at $r = 2^{1/6}\sigma$, so that the state functions from the previous section can be applied here by the transformation of $x = r - 2^{1/6}\sigma$. This system would have the selection rules for the harmonic oscillator \cite{Pilar}, with an energy difference between adjacent levels of $\epsilon/\gamma^2$ so that one could examine experimental data for the vibrational spectrum of the known diatomic and, from this, establish an estimate for $\gamma^2$.

\section{Conclusion:}
The one-dimensional quantum harmonic oscillator is written and solved in terms of parabolic cylinder function by employing a few simple substitutions. The solution provides a quicker, more direct method for solving the simple quantum harmonic oscillator and also for the harmonic oscillator within a uniform electric field. The energy eigenvalues, as well as the stationary states of both systems, can be evaluated with minimal computation. 

These results alert one to the fact that the oscillator in a uniform electric field can be considered for the cases where the field shifts the energy spectrum by an integer. From this, we proposed and gave a simple example of how the known solution for the oscillator in the uniform field model can be used to approximate the bound states of an L-J potential. Though we described the method of approximating an L-J potential here, this method could be used for any one-dimensional potential where the equilibrium position and minimum energy are known.

\end{document}